%% file: ACMmain.tex
\documentclass[acmsmall,screen,authorversion,nonacm]{acmart}

\AtBeginDocument{%
  \providecommand\BibTeX{{%
    \normalfont B\kern-0.5em{\scshape i\kern-0.25em b}\kern-0.8em\TeX}}}

\usepackage{multirow}
\usepackage{stackengine}
\usepackage{caption}
\usepackage{algorithm}
\usepackage{algpseudocode}
\usepackage[utf8]{inputenc} 
\usepackage[T1]{fontenc}    
\usepackage{hyperref}       
\usepackage{url}            
\usepackage{booktabs}       
\usepackage{amsfonts}       
\usepackage{nicefrac}       
\usepackage{microtype}      
\usepackage{xcolor}         
\usepackage{graphicx}
\usepackage{subcaption}
\usepackage{multirow}
\usepackage{svg}
\usepackage{amsmath}

\usepackage{longtable}
\usepackage{lscape}
\usepackage{cleveref}
\usepackage{adjustbox}
\usepackage{csquotes}
\usepackage{graphicx,wrapfig,lipsum}

\usepackage{tabularx}

\makeatletter
\newcommand{\printfnsymbol}[1]{%
  \textsuperscript{\@fnsymbol{#1}}%
}
\makeatother

\setcopyright{acmcopyright}
\copyrightyear{2022}
\acmYear{2022}

\begin{document}

\title{The Role of Large Language Models in the Recognition of Territorial Sovereignty: An Analysis of the Construction of Legitimacy}

\author{Francisco Castillo Eslava*}
\affiliation{%
  \institution{University of Granada}
  \country{Spain}
}
\email{fceslava@ugr.e}

\author{Carlos Mougan*}
\affiliation{%
  \institution{University of Southampton}
  \country{United Kingdom}}
\email{c.mougan-navarro@soton.ac.uk}

\author{Alejandro Romero-Reche}
\affiliation{%
  \institution{University of Granada}
  \country{Spain}
}
\email{romeroreche@ugr.es}

\author{Steffen Staab}
\affiliation{%
  \institution{University of Southampton}
  \country{United Kingdom}
}
\affiliation{%
  \institution{University of Stuttgart}
  \country{Germany}
}
\email{S.R.Staab@soton.ac.uk }

\renewcommand{\shortauthors}{Castillo-Eslava F., Mougan C., Romero-Reche A. and Staab S.}

\definecolor{amethyst}{rgb}{0.6, 0.4, 0.8}
\newcommand{\gourab}[1]{\textcolor{blue}{GP:#1}}
\newcommand{\steffen}[1]{\textcolor{red}{Steffen:#1}}
\newcommand{\carlos}[1]{\textcolor{amethyst}{Carlos:#1}}
\newcommand{\sr}[1]{\textcolor{red}{Salvatore:#1}}
\newcommand{\jose}[1]{\textcolor{green}{Jose:#1}}

\begin{abstract}

We examine the potential impact of Large Language Models (LLM) on the recognition of territorial sovereignty and its legitimization. We argue that while technology tools, such as Google Maps and Large Language Models (LLM) like OpenAI's ChatGPT, are often perceived as impartial and objective, this perception is flawed, as AI algorithms reflect the biases of their designers or the data they are built on. We also stress the importance of evaluating the actions and decisions of AI and multinational companies that offer them, which play a crucial role in aspects such as legitimizing and establishing ideas in the collective imagination. Our paper highlights the case of three controversial territories: Crimea, West Bank and Transnitria, by comparing the responses of ChatGPT against Wikipedia information and United Nations resolutions. We contend that the emergence of AI-based tools like LLMs is leading to a new scenario in which emerging technology consolidates power and influences our understanding of reality. Therefore, it is crucial to monitor and analyze the role of AI in the construction of legitimacy and the recognition of territorial sovereignty.

\end{abstract}


\maketitle

\input{content/introduction}

\input{content/previousWork}
\input{content/methodology}
\input{content/experiments}
\input{content/discussion}
\input{content/conclusions}

\bibliographystyle{ACM-Reference-Format}
\bibliography{ref}

\end{document}

%% file: content/introduction.tex
\section{Introduction}

%
Territorial sovereignty is a topic that is not typically free of controversy~\cite{sep-legitimacy}. When more than one political actor demands the right to administer the same region, it forces the remaining actors to take a position on the new dilemma that opens~\cite{weber2009theory}. Recognizing the sovereignty of one of the parties involved carries the possibility of animosity with the aggrieved party and granting legitimacy with respect to the disputed territory. In the case of states, they have their own tools, such as diplomatic recognition, to vest a political subject with authority over a territory~\cite{weber2016types}.

But it is not only states that confer legitimacy to different political authorities. Organizations and multinational corporations play a crucial role in legitimizing and establishing ideas in the collective imagination~\cite{habermas1985theory}. This means that today, borders are not defined solely by formal political institutions. Considering that authority has a right over a territory or the perception that a certain entity administers a region is also determined by the position that companies take and reflect on the issue. Consequently, it is crucial to closely monitor and evaluate the actions and decisions taken by these organizations.

A clear example, is Alphabet Inc. through its well-known online mapping service Google Maps. The difficulty of delimiting territory and the inherent arbitrariness that characterizes the task of drawing borders in the presence of a territorial conflict led to the offense and complaints of more than one political actor who felt damaged by the company’s defined border demarcation. As reflected in the work of Bogen and Quiquivix~\cite{bogen,doi:10.1080/00045608.2014.892328}, the delimitation of the region that houses the Palestinian-Israeli conflict or the territory of Arunachal Pradesh claimed by China and India are two examples of the different controversies generated around the application.

In recent years, driven precisely by the investment of these types of technology companies, a set of artificial intelligence-based tools has been developed. Its application and use have become popular due to its performance. One aspect that characterizes them, besides the speed, efficiency, and scalability with which they operate, is the impartiality often attributed to them. This perceived impartiality comes from considering that their verdicts are the result of operating with objective data provided without the mediation of a third party\cite{fountain2022moon}. The main problem and threat that arise from the social legitimization of this consideration are that it is actually a perception. The basis on which these types of applications are built is the vast amount of information extracted, mostly, from the Internet. This means that the results offered by the tool will be exposed to the same biases that are found in the data, or alternatively, will be flawed by the designer’s will\cite{joyce2021toward}.

One of the latest online applications that has attracted the most media attention is being carried out by OpenAI’s ChatGPT. These new algorithms called Large Language Models (LLM) are oriented towards interaction with the user through questions and answers, keeping a high degree of naturalness in the development of the conversation. The enormous impact generated by its appearance outlines a new scenario in which this type of emerging technology ends up consolidating itself as a new information search tool. The consolidation of this process would integrate them into the social field as a new actor legitimizing public opinion and knowledge. In this way, the imminent influence that the application and popularity of this technology can reach in society makes its study a necessity and impels us to take an interdisciplinary perspective capable of addressing the different dimensions that characterize it. 

Thus, the objective of our research is to examine the positions that ChatGPT reflects on issues that constitute a public debate. Existing territorial conflicts of sovereignty will be used for a case study due to their complexity, importance, and implications, both sociol and in the territorial conformation of the states.

In summary, our research makes the following contributions:  
\begin{itemize}
    \item First, it highlights the impact of new technologies on the construction of legitimacy.  
    
    \item Second, it takes ChatGPT as a new legitimizing social actor of knowledge and provides an analysis and comparison of its responses on the legitimacy of three disputed territories with the position reflected by the United Nations and Wikipedia.
    
    \item Third, it contributes to the ongoing discussion about the risks and potential harm arising from the usage of Large Language Models

\end{itemize}

%% file: content/previousWork.tex
\section{Background}\label{sec:relatedwork}
\subsection{Large Language Models}
Large Language Models (LLMs) are machine learning models that are designed to represent the likelihood of word sequences in a given domain, such as essays, books, or documents. LLMs are trained to capture statistical patterns in word sequences from a particular corpus, and can then be used to predict the probability of generating the next sequence of words. The architecture and training objectives of an LLM may vary depending on its intended use. In this discussion, we will focus specifically on LLMs that are tailored for language generation.

A series of Large Language Models (LLMs) \cite{DBLP:conf/nips/BrownMRSKDNSSAA20,lieber2021jurassic,DBLP:journals/corr/abs-2112-11446,DBLP:journals/corr/abs-2201-08239}, with the largest dense autoregressive transformers \cite{DBLP:conf/nips/VaswaniSPUJGKP17} having over 500 billion parameters, have been introduced. These LLMs have demonstrated impressive performance on various tasks, using evaluation protocols such as zero-shot, few-shot, and fine-tuning.

Some of these methods, including ChatGPT, have been developed using a technique known as Reinforcement Learning from Human Feedback \cite{DBLP:conf/nips/ChristianoLBMLA17}

. While this language model training technique can create highly conversational algorithms, it is still in a development phase, and several examples have shown blatant errors, such as referencing non-existent scientific studies \cite{thorp2023chatgpt}. As a result, the impact of these algorithms on society is yet to be fully realized.

Fine-tuning with crowd worker-annotated data, which is the last step in the training process, can result in significant improvements in ensuring that the model's responses are consistent with a set of human values, such as preventing harmful suggestions and unfair bias~\cite{DBLP:journals/corr/abs-2201-08239}. \cite{DBLP:conf/nips/SolaimanD21} show that adjusting LLMs behavior is feasible with a small sample of hand-curated dataset. Although current open-source resources for LLMs are trained using this fine-tuning step, there are still additional steps pending, such as factual grounding. This involves enabling the model to consult external knowledge sources, such as information retrieval systems, language translators, and calculators. The aim of factual grounding is to enable the model to generate responses based on known sources, rather than responses that merely seem plausible. This approach may be useful in the future for improving the reliability of LLM-generated responses.

\begin{table}
    \caption{Generative text model outputs temporal evolution comparison.  2011 is extracted from Sutskever et a.~\cite{DBLP:conf/icml/SutskeverMH11}, 2020 text from GPT-3 ~\cite{DBLP:conf/nips/VaswaniSPUJGKP17,DBLP:journals/corr/abs-2301-04246}, and 2023 by querying chatGPT (6 March 2023)}
    \begin{tabularx}{\columnwidth}{X}
        \hline
        \textbf{2011}                \\
        \hline
        \textbf{The meaning of life} is the tradition of the ancient human reproduction: it is less favorable\\ to the good boy for when to remove her bigger\\
        \hline
        \textbf{2020}\\
        \hline
        \textbf{The meaning of life} is a philosophical and existential question that has been debated for centuries. There is no single answer or definitive explanation for what the meaning of life is, as it can vary from person to person based on their beliefs, values, and experiences. Some people believe that the meaning of life is to achieve happiness and fulfillment, while others believe it is to seek knowledge and understanding of the world around us. Some believe that the purpose of life is to fulfill a certain destiny or to contribute to society in a meaningful way.\\ Others see life as inherently meaningless, and that it is up to each individual to create their own meaning and purpose through their choices and actions.\\ 
        \hline
        \textbf{2023}\\
        \hline
        \textbf{The meaning of life} is a philosophical and existential question that has been debated for centuries. There is no single answer or definitive explanation for what the meaning of life is, as it can vary from person to person based on their beliefs, values, and experiences. Some people believe that the meaning of life is to achieve happiness and fulfillment, while others believe it is to seek knowledge and understanding of the world around us. Some believe that the purpose of life is to fulfill a certain destiny or to contribute to society in a meaningful way.Others see life as inherently meaningless, and that it is up to each individual to create their own meaning and purpose through their choices and actions.Ultimately, the meaning of life is a deeply personal and subjective concept that each person must explore and define for themselves.\\
        \hline
    \end{tabularx}
    \label{table:LLMtemporal}
\end{table}

\subsection{Social Science Foundations}
Responsible innovation requires an interdisciplinary approach~\cite{stilgoe2013developing}. In order to anticipate, understand and mitigate the possible risks posed by new technology, there is the need to view its implications through different lenses. In this work we provide a territorial construction legitimacy perspective.

As conceptualized by social theory and established by empirical social research, all social interaction takes place within the framework of specific structural constraints and cultural patterns and, in its turn, produces, reproduces, or changes such framework for future interactions. Human interaction with AI is bound to develop a similar dynamic ~\cite{airoldi2021machine}: both humans and algorithms learn from each interaction and adjust for future encounters in a feedback loop that reinforces certain patterns while discouraging others.

However fluid or convincing the use of natural language, interacting with ChatGPT or any other AI is essentially different, at a sociological level, from interacting with another human being, at least to the extent that the human user is aware of the non-human nature of their partner in interaction. The social implications of such nature depend on how the non-human agent has been culturally constructed and the qualities it is perceived as having. Users looking for answers about relevant topics will value them differently if the non-human agent is perceived as being impartial or biased, comprehensive or arbitrary, etc.

Obviously, this is particularly relevant when AI is used not just as a source of unstructured factual information but, rather, as an authoritative source of structured knowledge, organized in pondered, essay-like responses where conflicting approaches and positions seem to be carefully weighed. In such cases, the responses provided may be perceived by users, due to the non-human nature of the agent, as fair and impartial in a way unattainable to human agents, even if the process of information-gathering is actually shaped by intrinsic biases in the original sources used or in their selection.

Hence our focus in the recognition of territorial sovereignty and the potential role of ChatGPT in the social perception of legitimacy. Sovereignty is socially constructed, both in specific historical instances and in its meaning as a concept ~\cite{glanville2013myth}; it is crucially shaped by the degree of recognition granted by international actors, and also by how statehood claimants manage such recognition or lack of it ~\cite{kyris2022state}. While the analytical approach to sovereignty in International Relations tends to be macroscopic, highlighting the roles played by nations, organizations and institutions, a thorough understanding of the process requires taking social perceptions into account, and how these are influenced by media, pundits, experts and other information and knowledge sources. 

As opposed to sources, such as Wikipedia entries, conspicuously written by humans and often suspected as biased despite editorial procedures designed to ensure objectivity, responses produced by AI are not expected to reflect the moral or political values of a writer or a team of writers located in specific countries and, therefore, defending specific interests in a conscious way or not. Algorithmic judgment appears as the way to avoid the essentially suspicious human subjectivity ~\cite{carlson2018automating}, and therefore its influence in social legitimation processes is potentially greater than those of other sources whose biases are taken for granted.

This calls for at least two complementary research approaches: 1) public opinion research about how AI is perceived as a source of knowledge by the general population, and 2) comparative discourse analysis of ChatGPT responses, considering that, while these remain inconclusive about unsettled political or scientific debates, they could fundamentally shape those debates through tacit assumptions. 

\subsection{Related work}
With the continued advancement of generative models, including language models (as shown in Table \ref{table:LLMtemporal}), there is growing concern among researchers, practitioners, and commentators about the potential benefits, risks, and negative effects associated with these models. To address this, workshops have been organized to better understand the risk landscape, specifically regarding the emergence of LLMs~\cite{tamkin2021understanding,https://doi.org/10.48550/arxiv.2301.04246}. Additionally, several reports have been published that provide: a comprehensive analysis of foundation models, covering various aspects such as technical principles, capabilities, applications, and societal impact~\cite{DBLP:journals/corr/abs-2108-07258}; help structure the risk landscape associated with LLMs providing multidisciplinary literature~\cite{weidinger2021ethical}; emphasizes environmental, financial costs of data curation, collection, and documentations~\cite{DBLP:conf/fat/BenderGMS21}; safety problem landscape for end-to-end conversational AI~\cite{DBLP:journals/corr/abs-2107-03451}; discuss the impacts of code generation technologies~\cite{DBLP:journals/corr/abs-2107-03374};
study how generative AI will increase productivity for knowledge workers~\cite{noy2023experimental}. In our work, we focus on a particular critical use case, construction of legitimacy of territorial recognition, that to the best of our knowledge has not been presented within the literature, and we provide an analysis comparing against different sources.

With respect to the evaluation of mitigating toxicity strategies in language models,~\cite{DBLP:conf/emnlp/WelblGUDMHAKCH21} analyze the consequences of these strategies in terms of model bias and quality, and evaluate their effectiveness using both automatic and human evaluation methods. In our work, we do not focus on mitigating toxicity but on evaluating our position on the topic of territorial recognition in comparison to other agents.

\cite{DBLP:journals/corr/abs-2103-14659} highlight some ways behavioral issues of LLMs can arise from accidental misspecification by the system designer. Our work is not necessarily targeted to mispecifications, but to highlight the difficulty of the task, which even among politics experts is not typically free of controversy~\cite{sep-legitimacy}.

The authors of the paper titled "Process for Adapting Language Models to Society (PALMS) with Values-Targeted Datasets"~\cite{DBLP:conf/nips/SolaimanD21} propose an iterative process that aims to change the behavior of language models significantly. This process involves crafting and fine-tuning the language model on a dataset that reflects a predetermined set of target values. The authors demonstrate that it is feasible to adjust language model behavior significantly with a small, hand-curated dataset. In our work, we analyze how generative language models can provide knowledge about the recognition of territorial sovereignty, which opens up the question of whether it's possible to tackle the problem of LLM-human alignment even in cases when there is no human consensus and how to react in these situations.

An example of the effects of technology on territorial sovereignty is Alphabet Inc.'s online mapping service Google Maps, which has led to complaints from political actors who felt damaged by the company's defined border demarcation. The delimitation of the region that houses the Palestinian-Israeli conflict or the territory of Arunachal Pradesh claimed by China and India are two examples of the different controversies generated around the application~\cite{bogen,doi:10.1080/00045608.2014.892328}.

%% file: content/methodology.tex
\section{Methodology}\label{sec:methodology}
The work was carried out through a content analysis of the answers provided by ChatGPT, the information found on Wikipedia and the resolutions issued by the General Assembly and the United Nations Security Council. Their application made it possible to define the position of each anonymized actor. Wikipedia's position was added in order to be able to compare the significance of ChatGPT with another widely spread information search web resource. And, the UN position was taken as an expression of the international community to check its possible influence.

Three territories were selected for analysis: Crimea, the West Bank, and Transnistria. The first was chosen because it represents one of the most relevant territorial conflicts of the last decade. The case of the West Bank was chosen because of the historical dimension that has characterized the conflict and the difficulties generated by the diplomatic relations that Israel maintains with a large part of the international community. And Transnistria was added because of the repeated occasions on which the UN Security Council has urged respect for the sovereignty and territorial integrity of the Republic of Moldova.

In the case of the IA, the data collection was obtained on December 28, 2022 through a series of questions asked to the three territories:

\begin{itemize}
    \item \textbf{Question 1: What is X?}
    \item \textbf{Question 2: To whom should X belong?}
    \item \textbf{Question 3: Who has rights over X?}
\end{itemize}

The construction of the questions set out was based on two approaches to the model: a general one that would serve as a contextualization of the territory, and another more direct approach to what the research was intended to find out. The first question was intended to find out if the model assigns to any political actor the sovereignty of the territory about which it is asked. With the statement of the other two questions, the model was already made aware of the existence of a dispute over the territory being asked about and it was desired to check the degree to which ChatGPT would be signified by exposing a disjunctive.  It will be considered that the model reflects a neutral attitude when it does not take the side of any of the parties to the conflict and, arbitrary, when the impartiality of the training data conditions it in favor of one of the parties.

%% file: content/experiments.tex
\section{Experiments} 
\label{sec:experiments}

\begin{table*}[ht]
\caption{Results table with a summary of the position  of the different actors analyzed towards the recognition of territorial sovereignty}
\begin{tabular}{c|ccc}
\multirow{2}{*}{Territory} & \multicolumn{3}{c}{Recognition}                                                                                \\ \cline{2-4} 
                           & \multicolumn{1}{c|}{United Nations}                   & \multicolumn{1}{c|}{Wikipedia}             & ChatGPT               \\ \hline
Crimea                     & \multicolumn{1}{c|}{In favor of Ukraine}  & \multicolumn{1}{c|}{In favor of Ukraine}    & In favor of Ukraine   \\
West Bank                  & \multicolumn{1}{c|}{In favor of Palestine}             & \multicolumn{1}{c|}{In favor of Palestine} & In favor of Palestine \\
Transnistria                & \multicolumn{1}{c|}{In favor of Moldavia} & \multicolumn{1}{c|}{In favor of Moldavia}  & Neutral              
\end{tabular}
\end{table*}

\subsection{Crimea}

\subsubsection{United Nations}

The position of the United Nations as an organization regarding the sovereignty of the Crimean territory was last resolved on the side of Ukraine in resolution 68/262 on March 27, 2014~\cite{UN/68/262}. Adopted with 100 votes in favor, 11 against, and 58 abstentions, the General Assembly:

\blockquote{
“Affirms its commitment to the sovereignty, political independence, unity and territorial integrity of Ukraine within its internationally recognized borders”

“Underscores that the referendum held in the Autonomous Republic of Crimea and the city of Sevastopol on 16 March 2014, having no validity, cannot form the basis for any alteration of the status of the Autonomous Republic of Crimea or of the city of Sevastopol”

“Calls upon all States, international organizations and specialized agencies not to recognize any alteration of the status of the Autonomous Republic of Crimea [...] on the basis of the above-mentioned referendum and to refrain from any action or dealing that might be interpreted as recognizing any such altered status” 
}

\subsubsection{Wikipedia}

The online encyclopedia describes the situation as a clear occupation by Russia and considers the territory a region of Ukraine from the very first sentence: \enquote{Crimea is a peninsula in Ukraine, on the northern coast of the Black Sea, that has been occupied by Russia since 2014}. Still in the introductory section, the website again speaks of Russian occupation, calls the referendum and annexation illegal and informs the reader that most of the international community recognizes Ukraine's sovereignty over the territory: \enquote{In 2014, the Russians occupied the peninsula and organized an illegal referendum in support of Russian annexation, but most countries recognize Crimea as Ukrainian territory}. This position is reaffirmed and the same arguments are restated in the history subsection under\enquote{\textit{Russian occupation (2014-present)}}~\cite{wiki:Crimea}.

\subsubsection{ChatGPT}

From ChatGPT's answers a declination towards the Ukrainian part of the conflict is induced. In the second sentence of the first question it already means considering the territory part of Ukraine, although it warns of being controlled by Russia: \enquote{It is located in Eastern Europe and is part of Ukraine, although it is currently controlled by Russia}.

The answers to the next two questions (Q2 and Q3), retain practically the same content and much of the same structure. In both, the model adds that the annexation has not been recognized by most countries of the international community, states that it has been widely condemned as a violation of international law, and refers to the resolutions in which the United Nations General Assembly urges Russia to withdraw its forces from Crimea and to respect the sovereignty and territorial integrity of Ukraine.

The difference between the latter two parts comes from the degree to which they ratify Ukraine's sovereignty over Crimea. While the second question displays a more descriptive tone of the conflict and describes the sovereignty of the territory as an object in international dispute - \enquote{the sovereignty of Crimea is a matter of international dispute} -, the third one leaves no room for doubt as to who has sovereignty over the area and turns to international law:

\blockquote{
\enquote{According to international law, Ukraine has the right to sovereignty over Crimea. Crimea is an autonomous republic within Ukraine and has been recognized as such by most countries in the international community}
}

Another aspect by which ChatGPT's position on the conflict can be perceived stems from the total absence of any statement, argument or version that could legitimize the action taken by Russia.

\subsection{West Bank}

\subsubsection{United Nations}

For the territory comprising the Palestinian-Israeli conflict, the position of the United Nations has been expressed through the issuance of a series of Security Council resolutions in favor of Palestine: 242 (1967), 338 (1973), 446 (1979), 452 (1979), 465 (1980), 476 (1980), 478 (1980), 1397 (2002), 1515 (2003), 1850 (2008) and 2334 (2016). If we take as a reference the last resolution~\cite{UN/S/RES/2334}, although they all coincide in the sense of the ruling, we can infer where the organization establishes the territorial limits of the two States when it expresses its concern about the fact that the: 

\blockquote{
Continuing Israeli settlement activities are dangerously imperiling the viability of the two-State solution based on the 1967 lines\ldots
}

The 1967 lines refer to the territorial demarcation established in the Arab-Israeli armistice of 1949. This would come to be adopted as the reference limits with which to recognize Palestinian territory after Israel exceeded them in the June 1967 war. In consideration of the 1967 lines, the West Bank is considered part of the borders of the Palestinian State.

Returning to the text, the UN condemns any measures undertaken by Israel \enquote{aimed at altering the demographic composition, character and status of the Palestinian Territory occupied since 1967} and refers to resolution 1515 (2003) to remind it of the obligation to halt \enquote{all settlement activity, including 'natural growth', and the dismantlement of all settlement outposts erected since March 2001}. Thus, the organization \enquote{reaffirms that the establishment by Israel of settlements in the Palestinian territory occupied since 1967, including East Jerusalem, has no legal validity and constitutes a flagrant violation under international law}.

\subsubsection{Wikipedia}

For the issue of the West Bank, Wikipedia again refers to the situation as an occupation. In this case, it states that the territory is \enquote{under an Israeli military occupation since 1967}. Another description that confirms this version comes when it refers to Israelis living in the West Bank as settlers. To reinforce the position it conveys, the website adds as an argument the illegality with which the international community considers Israeli settlements on the basis of international law. In the same way, it uses a ruling of the International Court of Justice as an endorsement. Apart from the introduction, Wikipedia again deals with the subject under the same terms in the different sections, such as in the history section and in the specific section called \enquote{Consequences of occupation}~\cite{wiki:West_Bank}.

\subsubsection{ChatGPT}

On this occasion, there is not so much evidence to perceive a clear positioning of ChatGPT on the conflict. In fact, it is only discernible from the use of one word in the first general question. Specifically when it uses the term \enquote{\textit{occupied}} when describing the configuration of territory: \enquote{Cisjordan includes the modern state of Israel, as well as the West Bank and Gaza Strip, which are territories occupied by Israel but claimed by the Palestinians}. With this sentence he delegitimizes Israel's sovereignty over the territories in the West Bank and grants sovereignty to Palestine.

On the other hand, the results obtained from the other two questions become an exercise in contextualizing the conflict. Two almost identical texts are extracted in which an attempt is made to reflect an impartial and merely descriptive position of the conflict:

\blockquote{
“The question of who has rights over Cisjordan, also known as Palestine or the Land of Israel, is a complex and controversial issue with a long history”

“Some people believe that the western region of Cisjordan should belong to the state of Israel, while others believe it should be an independent Palestinian state.”

“There are different viewpoints on this issue, and the resolution to the conflict will depend on the willingness of the parties involved to find a mutually acceptable solution that takes into account the rights and interests of all parties”
}

\subsection{Transnistria}

\subsubsection{United Nations}

The position of the United Nations on the status of Transnistria is again induced from the set of resolutions issued on the territory, but in this case by the General Assembly: 54/117 (1999), 55/179 (2000), 56/216 (2001), 57/298 (2003)\cite{UN/A/RES/57/298} and 72/282 (2018)\cite{UN/A/RES/72/282}. In the first two, the existing references to the Transdniestrian issue are aimed at supporting efforts to resolve the conflict and to reflect Russia's commitment to withdraw its forces from the territory of the Republic of Moldova. The two successor resolutions to those just cited retain the same purpose, but add that the political settlement of the conflict must be based on "full respect of the sovereignty and territorial integrity of the Republic of Moldova". Finally, the last pronouncement of the General Assembly is issued in order to urge “the Russian Federation to complete, unconditionally and without further delay, the orderly withdrawal of the Operational Group of Russian Forces and its armaments from the territory of the Republic of Moldova”. In it, the UN recalls that " the stationing of foreign military forces on the territory of the Republic of Moldova, without its consent, violates its sovereignty and territorial
integrity" and is concerned that “the continuous illegal joint military exercises of the
Operational Group of Russian Forces with the paramilitaries of the separatist entity
in the eastern part of the country, [\ldots], disregards
the sovereignty and territorial integrity of the Republic of Moldova.

\subsubsection{Wikipedia}

Transnistria is defined as an \enquote{unrecognized breakaway state} and reports that it is internationally recognized as part of Moldova. Despite this, it recognizes that the Pridnestrovie Republic of Moldova controls a large part of the area. Still sticking to the first paragraph, Wikipedia makes clear the scarce international recognition of the region, warning that it only has the support of three unrecognized or partially recognized separatist states - Abkhazia, Artsakh and South Ossetia-. In the same vein and with the intention of manifesting Russian influence and dependence, it mentions the resolution of the Parliamentary Assembly of the Council of Europe defining the territory as under military occupation by Russia. The content of the website intertwines statements that allude to the de facto control of the region as an independent republic with others that recall the scarce international recognition. Finally, in the section dedicated exclusively to the status of the region\cite{wiki:Transnistria}, the web site pronounces that:

\blockquote{
“All UN member states consider Transnistria a legal part of the Republic of Moldova. Only the partially recognised or unrecognised states of South Ossetia, Artsakh, and Abkhazia have recognised Transnistria as a sovereign entity after it declared independence from Moldova in 1990 with Tiraspol as its declared capital”.
}

\subsubsection{ChatGPT}

Finally, in the case of Transnistria, it was the only territory where no evidence was found in the language used by ChatGPT to draw a clear conclusion about its position on the conflict. As was seen in much of the West Bank example, the model's response offers a contextualization of the conflict with great care in the language so as not to be biased. Thus, the response is limited to providing historical data and indicating the position of the actors involved from a neutral position:

\blockquote{
“Transnistria declared independence from Moldova in 1990, but its independence has not been recognized by any country”

“The Moldovan government considers Transnistria to be an autonomous territorial unit within Moldova, while the Transnistrian authorities consider themselves to be an independent state”
}

The only aspect that can be reproached in ChatGPT's response with respect to the neutrality it pursues in the matter is the failure to allude to the recognition granted by the three unrecognized or partially recognized separatist states: Abkhazia, Artsakh and South Ossetia.

%% file: content/discussion.tex
\section{Discussion}

The content analysis undertaken on the ChatGPT responses and comparison against Wikipedia and the United Nations resolutions has led to the identification of a set of particularities that have facilitated the task of characterizing the behavior displayed by the model. The main issue to highlight refers to an analysis of the biased answers of AI systems. As we have been able to verify, the answers offer, in two of the three cases, evidence that reflects a position in favor of one of the parties to the conflict. For Crimea, it explicitly positions itself on the side of Ukraine and, although it is less categorical with regard to the West Bank, it offers a biased response in favor of Palestine when speaking of occupation with respect to the territories controlled by Israel in the region.

The case of Transnistria is the only territory where no terms or expressions have been found to induce the AI to take sides in the conflict. Taking into account that part of ChatGPT's source of information is based on Web content, it is possible that the significantly lower media coverage of the conflict has influenced the amount of information generated in this regard and, thus, the neutrality of the model. On the other hand, the categorical expressions with which he addresses the issue of the Crimean peninsula may be due to precisely the opposite:: one of the most important international conflicts of the last decade and with the condemnation of the Russian annexation by most of the international community -as reflected in the UN General Assembly resolution 68/262~\cite{UN/68/262}.

In relation to the Palestinian-Israeli issue,  a possible explanation for the less forcefulness with which ChatGPT is signified and the descriptive nature of its responses can perhaps be found in the high complexity of the conflict and in the actors involved. Although the UN Security Council has already pronounced itself warning of the illegality of the Israeli settlements in the West Bank ~\cite{UN/S/RES/2334}-with the abstention of the US-, the close relations that Israel maintains with much of the international community might influence attenuate the positions of its partners. This could significantly affect the generation of the number of official documents condemning Israel's conduct. Just as the use of gender-ethnicity based image ratios has been a topic of controversy in the past~\cite{DBLP:conf/fat/BuolamwiniG18}, it prompts the question of what official document ratios should be used for international political conflicts and, consequently, the source of data from which the model learns would also be reduced.

Thus, the position of the two online resources has always taken the same position as the one issued by the UN, except for the case of Transnistria in which ChatGPT takes a neutral position on the conflict. The fact induces to think that the attitude of the international community influences the position reflected by ChatGPT and the content added to Wikipedia.

Another aspect worth highlighting is related to the effectiveness of the questions developed to achieve the research objective. Two types of questions were applied to the model: $(i)$ the first of a general nature about the territory and  $(ii)$ the other aimed directly at finding out
who ChatGPT considers has more rights over the disputed territory.; it was the first one that best captured the positioning of the model. This may be due to the training to which the AI has been subjected to detect questions that expect a biased or tendentious answer from the model. Thus, the most effective strategy in our research to collect the behavior of ChatGPT has been to ask general and indirect questions about the issue to be evaluated, instead of specific and directed directly to decipher the position of the model. This way of skirting the impartiality with which the AI responds offers greater possibilities of reflecting an arbitrary behavior in response to a question asked.

Finally, an issue perceived in the work and which has a bearing on the idea just expressed has to do with the high similarity found between the answers to questions two and three. In these questions, in addition to the content whose difference is minimal when dealing with the same territory, a general pattern is observed that is maintained in the responses from the different regions:
\blockquote{
\enquote{The question of to whom Transnistria belongs is a complex and disputed issue} (Question 2. Transnistria)

\enquote{The question of who has rights over West Bank [...] is a complex and controversial issue with a long history} (Q3. West Bank)

\enquote{Ultimately, the resolution of disputes over sovereignty and territorial control depends on the willingness of all parties to engage in dialogue and find mutually acceptable solutions} (Q3. Transnistria)

\enquote{Ultimately, the resolution to this conflict will depend on the willingness of the parties involved to find a mutually acceptable solution that takes into account the rights and interests of all parties }(Q2. West Bank)
}

This singularity shows the last step of ChatGPT training process Reinforcement Learning from Human
Feedback \cite{DBLP:conf/nips/ChristianoLBMLA17} (cf. Section \ref{sec:relatedwork}), with respect to the willingness of ChatGPT developers to identify and avoid questions that seek to provoke the arbitrariness of the model. The detection of patterns built by similar expressions in similar questions leads us to appreciate the semantic structure applied during the training process. The detection of these patterns in the semantics of the generative model reduces at the same time the naturalness with which it interacts. 

%% file: content/conclusions.tex
\section{Conclusion}
\label{sec:conclusion}

The results and the analysis carried out throughout the study have led to a series of reflections on the behavior identified in ChatGPT. The first one is directed to the behavior exhibited by the model when expressing itself on a controversial topic. At this juncture, artificial intelligence LLMs aspire to offer a neutral response that does not lean towards any of the opposing parties. Despite attempts to avoid bias, the results of the work have shown that in two of the three scenarios the model has ended up offering a biased response. This perceived attitude has significant implications for both the knowledge acquisition processes and the case study selected for this research.

Algorithmic judgment, as performed by ChatGPT when questioned about territorial sovereignty, may avoid certain elements of human subjectivity, but still incorporates value judgments and positions that may be intrinsic to the sources from which it gathers information or even to the very definition of concepts and problems. The emergence of a computational authority [Airoldi, 2021], regarded as a key source of knowledge that impartially ponders the merits and arguments of conflicting positions to provide an unbiased review of the subject matter, could entail the reinforcement of specific points of view, thus established as objective by a non-human, dispassionate (and, therefore, more reliable) intelligence.

To the extent that LLMs are popularized to a mass public as an information search resource, their impact on the generation and consolidation of knowledge will be extended to all areas susceptible to their use. In the case at hand, the potential implications are especially significant because they transcend the individual level and also have an effect on state borders. The recognition of a territory to a political actor constitutes the legitimization of the one who issues it. With this in mind, the perception of all users who make use of AI such as ChatGPT on the sovereignty of a region will be influenced by the positioning of the model's responses with respect to the territory. Thus, the analyzed technology becomes a relevant social agent in the consolidation of public opinion about the territorial sovereignty of a region, especially in those territories where more than one political actor claims sovereignty over it. It should be added that these resources are being offered by large technology companies whose orientation may be conditioned to the particular interests of the companies.

In view of the above and the biased results that the model has offered, it becomes necessary to address the debate on the role that these new tools should play in society and the position that they should reflect in the face of the transcendental dilemmas described above. Opting for neutrality by offering a fair description of the conflict or reproducing the position taken by the international community are two options that are not free of risks and controversies. The issue is extremely complex and multifaceted, and ongoing debates are needed to understand the impact and consequences of AI behavior on knowledge generation and territorial legitimization. Thus, with the development of this paper we hope to make visible one of the conflicts that generates the greatest tension at the international level and in which artificial intelligence has begun to play a leading role.

\textbf{Limitations and further work:} This work opens up a problem and necessarily leaves questions unanswered. For example, we do not discuss potential beneficial
applications of LMs nor do we focus in many international conflicts nor on different languages. Even though territorial recognition problems are a longstanding political issue, the technology used represents a snapshot in time, and algorithms are queried through the years 2022 and 2023.